 \documentclass[doublespacing]{elsart}

 \usepackage{graphics}
 \usepackage{graphicx}
\usepackage{amssymb}

\begin{document}

\begin{frontmatter}

\title{Light ions response of silicon carbide detectors\thanksref{4}}

\author[1]{M. De Napoli},\author[1]{G. Raciti\corauthref{3}},\ead{raciti@lns.infn.it}\author[1]{E. Rapisarda},
\author[2]{C. Sfienti}\\

%%%%%%%%%%%%%%%%%%%%%%%%%%%%%%%%%%%%%%%%%%%%%%%%%%%%%%%%

\address[1]{Dipartimento di Fisica, Universit$\grave{a}$ degli studi di Catania and Sezione INFN \\ Via S.Sofia,64 I-95123
Catania-Italy}
\address[2]{GSI-Darmstadt-D-64291-Germany}

%%%%%%%%%%%%%%

\corauth[3]{Corresponding author: Tel:+ 39-095-542 286 FAX: +
39-095-378 5225}
\thanks[4]{Partly supported by the European Community in the framework of the
"DIRAC secondary-beams" project, contract N.515873 under the
"Structuring the European Research Area" Specific Programme Research
Infrastructures Action.}

%%%%%%%%%%%%%%%%%%%%

\begin{abstract}
Silicon carbide (SiC) Schottky diodes 21 $\mu$m  thick with small
surfaces and high N-dopant concentration have been used to detect
alpha particles and low energy light ions. In particular $^{12}$C
and $^{16}$O beams at incident energies between 5 and 18 MeV were
used. The diode active-region depletion-thickness, the linearity of
the response, energy resolution and signal rise-time were measured
for different values of the applied reverse bias. Moreover the
radiation damage on SiC diodes irradiated with 53 MeV $^{16}$O beam
has been explored. The data show that SiC material is radiation
harder than silicon but at least one order of magnitude less hard
than epitaxial silicon diodes. An inversion in the signal was found
at a fluence of 10$^{15}$ ions/cm$^2$.
\end{abstract}

\end{frontmatter}

%%%%%%%%%%%%%%%%%%%%%%%%%%%%%%%%%%%%%%%%%%%%%%%%%%%%%%%%
% Write the text starting from here and using the usual
% LaTeX commands.
%~

\runtitle{Light ions response of silicon carbide detectors}
\runauthor{M. De Napoli, G. Raciti,E. Rapisarda,C. Sfienti}
\maketitle
\section{Introduction}
In the last years the use of electronic devices and sensors in
very harsh environments at elevated temperatures, high-power,
high-frequency and high radiation fields has became the subject of
research and development in various fields. In particular the
search for new materials, more suitable for such extreme operating
conditions than the usual silicon semiconductor has received a lot
of interest. Among the investigated materials, the silicon carbide
(SiC) semiconductor has raised large interest and has been already
used in a wide range of applications
\cite{Tre91,Bal94,Wei96,Hey98,Iee99,coo90,Ela02,Cla02}. In
particular, recent work has been done on the development of SiC
radiation detectors \cite{Rog99,Bru01} and on the characterization
of their performances. SiC detectors were used with excellent
results as neutron \cite{Dulo99} and X-ray detectors operating at
high temperatures \cite{Ber04}. The charged-particle response
characteristics have been measured by irradiating the detectors
with alpha particle sources at energies up to 5.48 MeV
\cite{Nav99,Rud03,Kino05}, and radiation damage effects were
investigated with 24 GeV protons and gamma rays
\cite{Cun02,Nav03,Cun04}, 300 MeV/c pions \cite{Cun03}, neutrons
\cite{Sci05} and protons, alphas and $^{12}$C beams \cite{Lee03}
at fluences up to 10$^{16}$ particles/cm$^2$ \cite{Sci05}
obtaining promising results. Indeed one of the most appealing
property of the SiC, as well as of other wide bandgap materials
such as GaAs and diamond, is their predicted radiation hardness
with respect to Silicon. Moreover SiC diodes show low reverse
current even at the very high electrical voltage applicable
because of their higher breakdown voltage with respect to Si. Both
properties make SiC diodes suitable for the construction of
specific detector arrays. The inner tracking detectors, typically
used in the high-radiation environment of particle physics
experiments, particle and ion detector arrays operating in
satellites or in the very harsh environment of laboratories
producing radioactive ion-beams would particularly benefit from
the mentioned properties. In spite of the notable development in
growing, processing and producing good-quality and low-defect SiC
diodes, work to date on the use of such diodes as detectors of
particles and ions has been very limited
\cite{Nav99,Rud03,Kino05,Cun03,Lee03}. We have therefore studied
the response signal of high-quality 4H-SiC Schottky diode to
$^{12}$C and $^{16}$O ions and to an alpha particles source at
various incident energies, in order to investigate the use of SiC
detectors also in nuclear physics applications. The main
objectives of the present work are the characterization of the
signal response in term of linearity, energy resolution and
rise-time as a function of the applied reverse bias and the
investigation of its degradation subsequent to irradiation with
light ions.

\section{Experimental details}
\begin{figure}[t]
\centering{\resizebox{0.6\textwidth}{!}{
\includegraphics{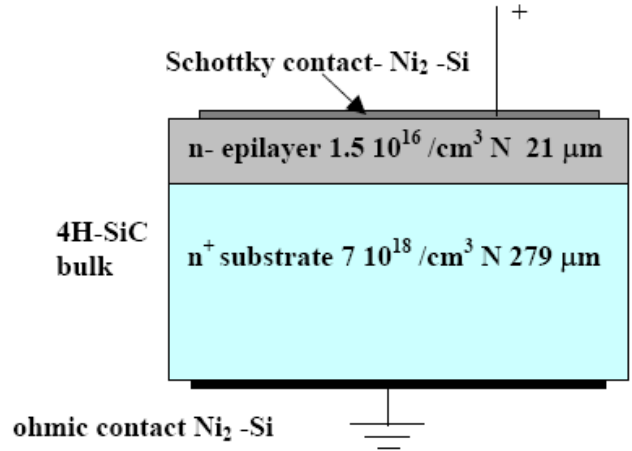}}
\caption{The 4H-SiC diode layout. The Schottky junction is realized
by the Ni$_2$Si layer, 0.2 $\mu$m thick, deposited at 600$^{\circ}$C
on the front surface, the ohmic layer by Ni$_2$Si deposited at
950$^\circ$C.}\label{fig:01}}
\end{figure}

\begin{figure}[h]
\centering{\resizebox{0.6\textwidth}{!}{
\includegraphics{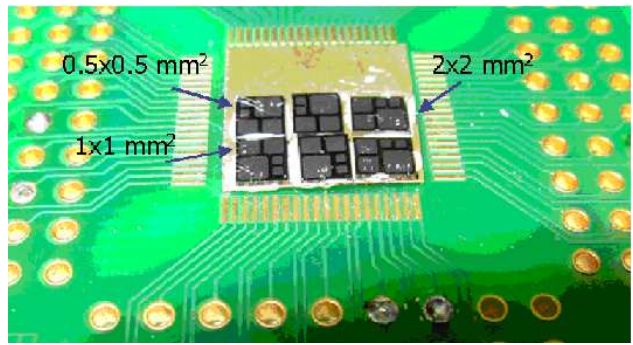}}}
\caption{Picture of the SiC detectors assembled on a board.}
\label{fig:02}
\end{figure}

The Schottky diodes were fabricated by epitaxy onto high-purity
4H-SiC n-type substrate from the ETC-Catania \cite{Etc06}.
Figure~\ref{fig:01} shows the layout of the detector. The nominal
 n$^-$ epitaxial layer nitrogen dopant concentration and thickness were,
respectively, 1.5$\times$10$^{16}$ N cm$^{-3}$ and 21 $\mu$m. The
dopant concentration and thickness of the n$^+$ side were
respectively 7$\times$10$^{18}$ N cm$^{-3}$ and 279 $\mu$m. The net
doping and the thickness have been measured \cite{Etc06},
respectively, by C-V (Hg-probe) and FTR Biorad QS-500. The nominal
defects mean value was 43 cm$^{-3}$ as measured \cite{Etc06} by
$\mu$Raman spectroscopy with an Ar-laser at 514.5 nm.\\ The Schottky
junction was realized by a 0.2 $\mu$m thick layer of Ni$_{2}$Si
deposited at 600 $^\circ$C on the front surface, while the ohmic
contact, on the back surface, was obtained with Ni$_{2}$Si deposited
at 950$^{\circ}$C \cite{LaV02}.\\
The active areas of the different SiC detectors were of 0.5x0.5,
1x1 and 2x2 mm$^2$ (see Fig.~\ref{fig:02}). The chips were glued
on a brass foil 1 mm thick by a conductive glue and single
contacts between the Ni$_{2}$Si front surfaces and the pads of the
board shown in Figure~\ref{fig:02} were realized by Al wire
(2$\mu$m thick) bonding.

%%%%%%%%%%%%%%%%%%%%%%%%%%%%%%%%%%%%%%%%%%%%%%%
Two boards with respectively six and five chips 2x2 mm$^2$ of
surface were assembled, for a total number of eleven independent
detectors. The third board was equipped with a Si detector 300
$\mu$m thick and 3x3 cm$^{2}$ of surface which was used as a
reference. The three boards were set-up in a scattering chamber at
the Laboratori Nazionali del Sud (LNS-Catania) and operated under
vacuum at 10$^{-6}$ mbar. The active samples were irradiated with
both 5.48 MeV alpha particles from an $^{241}$Am source and $^{12}$C
and $^{16}$O beams accelerated by the LNS Tandem VdG at different
energies and currents. The alpha source was set-up at a distance of
10 cm from the detector surface in order to get a near normal
incidence of alpha particles on the detector. The ion beams were
focused 60 cm downstream with respect to the detector position in
order to achieve an uniform perpendicular irradiation in a spot of
about 3 mm of diameter on the SiC surface. The boards holder was
moved so as to center the detectors one by one. Standard electronics
was used to process the signals: preamplifiers of 45 mV/MeV gain and
amplifiers with 0.5 $\mu$sec shaping time. Data acquisition was
based on CAMAC ADCs read out through a GPIB standard National
Instruments interface and a data-acquisition program built in the
LabView 7 framework. Moreover the preamplifier and amplifier signals
were digitalized by a Tektronix TDS 5104B digital oscilloscope.

\section{Detector response}

\begin{figure}[h]
\centering{\resizebox{0.6\textwidth}{!}{
\includegraphics{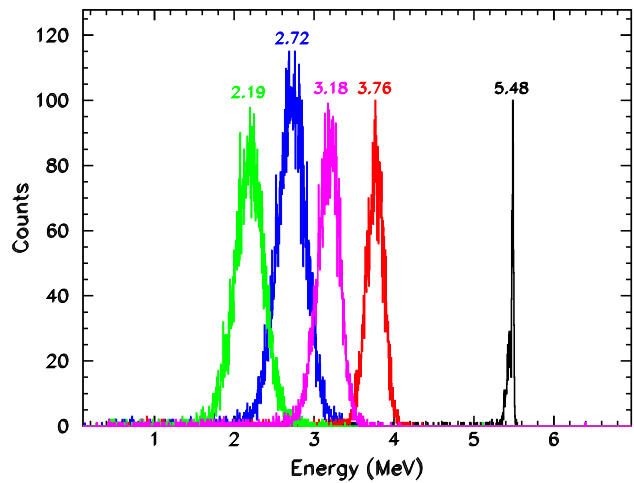}}}
\caption{The four $^{241}$Am alpha energy spectra, measured by the
Si detector with the 10, 13, 14.6 and 17.3 $\mu$m Al absorbers in
front of the source. The 5.48 MeV peak is also shown for
reference. }\label{fig03}
\end{figure}

\begin{figure}[th]
\centering{\resizebox{0.6\textwidth}{!}{
\includegraphics{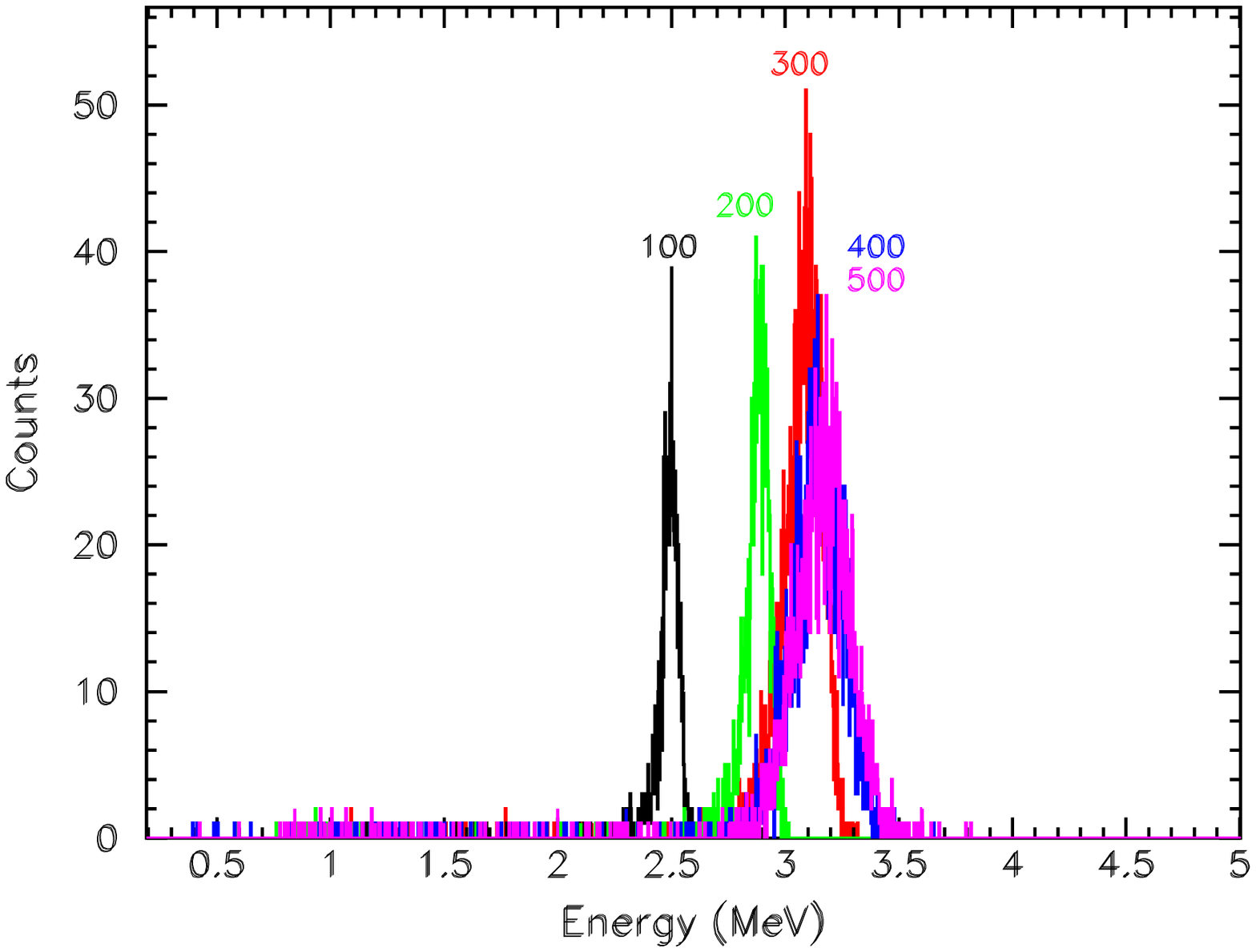}}
\resizebox{0.6\textwidth}{!}{
\includegraphics{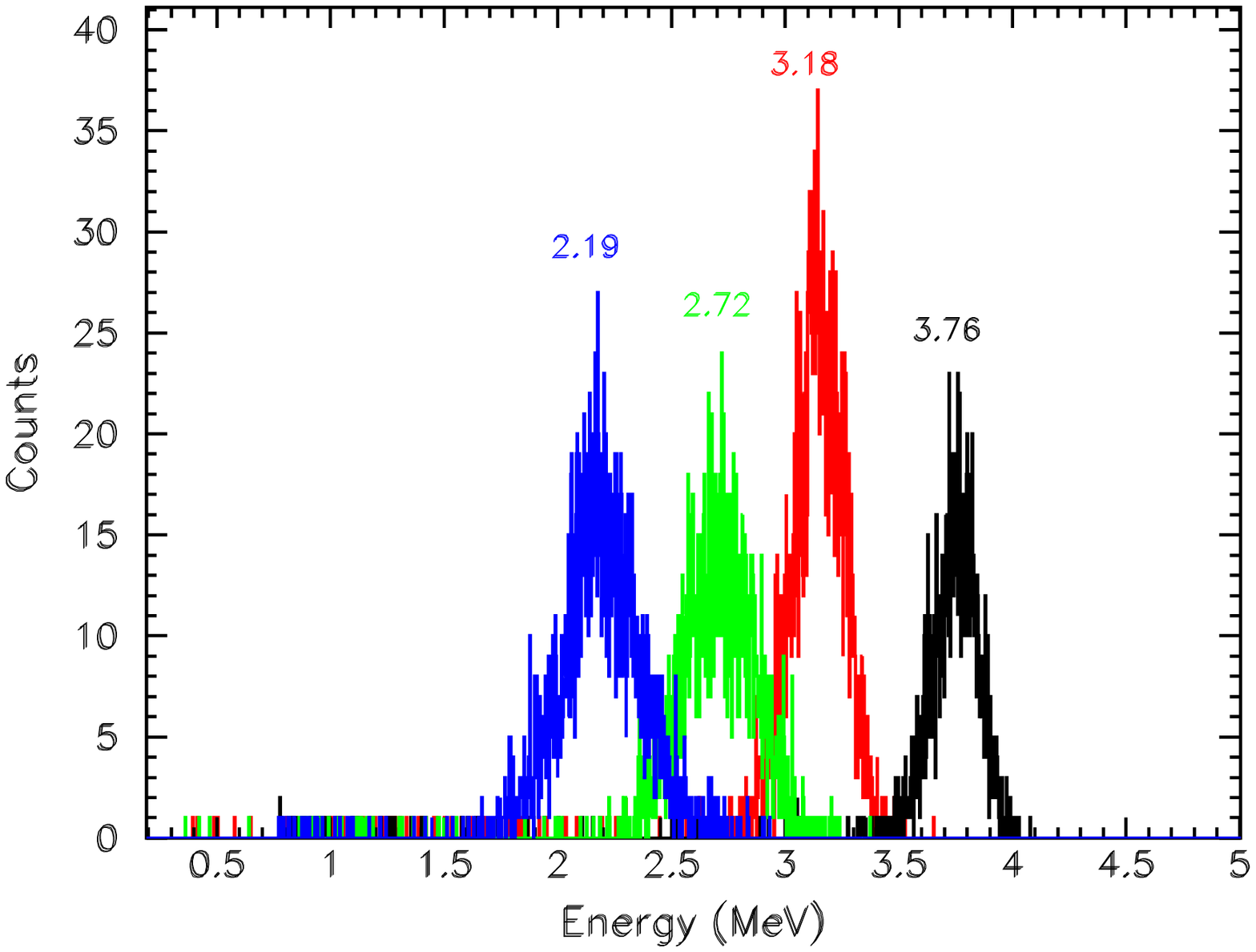}}}
\caption{Top: Energy spectra measured by the SiC diode at increasing
bias voltage for alpha particles of 3.18 MeV incident energy.
Bottom: The four alpha energy spectra (2.19, 2.72, 3.18, 3.76 MeV)
measured by the SiC diode at the bias saturation values.}
\label{fig04}
\end{figure}

\begin{figure}[h]
\centering{\resizebox{0.6\textwidth}{!}{
\includegraphics{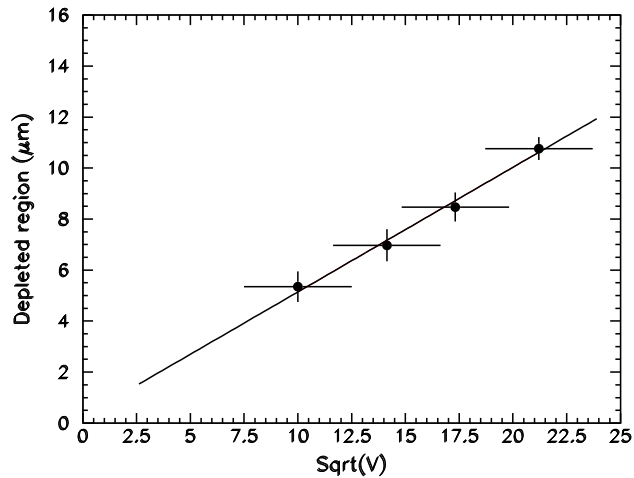}}}
\caption{Correlation between the square root of the applied bias and
the SiC depletion layer thickness evaluated from alpha energies.}
\label{fig05}
\end{figure}

It is well known that a reverse bias must be applied in order to
create in the diode a depleted region which acts as the active
region for the detection of the charges produced by
ionization from the incoming charged particles.\\
The order of magnitude of this reverse bias could be a relevant
parameter for specific applications. Therefore we measured at first
the correlation between the thickness of the depleted layer and the
applied bias, which is expected to follow a square-root law
\cite{Kno00}. By using four Al absorbers of 10, 13, 14.6 and 17.3
$\mu$m thicknesses, in front of the $^{241}$Am alpha source, we
scaled the incident energy from 5.48 MeV to 3.76, 3.18, 2.72 and
2.19 MeV respectively as measured by the calibrated silicon detector
(see Fig.~\ref{fig03}).
%%%%%%%%%%%%%%%%%%%%%%%%%%%%%%%%%%%%
Using the SRIM code \cite{Zie96} calculations, the ranges of the
alpha particles at the four incident energies in the SiC material
were evaluated to be: 10.8$\pm$0.4, 8.5$\pm$0.5, 7.0$\pm$0.6 and
5.4$\pm$0.5 $\mu$m respectively. The errors on the range thickness
were evaluated by taking into account both the energy loss in the
Ni$_2$Si 0.2 $\mu$m thick front layer on the detector and the FWHM
(full-width at half-maximum) of the alpha spectra of
Fig.~\ref{fig03}, in order to account for the energy straggling
introduced by the Al absorbers. A typical set of spectra is shown in
Fig.~\ref{fig04} for one of the detectors used. For all the
detectors the centroid position of the peak moves toward higher
channels as the voltage increases. The observed shift can be
described according to the increasing thickness of the active volume
of the detector with increasing reverse bias voltage.
%%%%%%%%%%%%%%%%%%%%%%%%%%%%%%%%%%%%
%%%%%%%%%%%%%%%%%%%%%%%%%%%%%%%%%%%%
As a result, the incident alphas deposit more energy in the active
region leading to detector signals with higher pulse heights. The
saturation of the pulse height values is then reached when the
applied reverse voltage depletes the active volume of the diode up
to the range corresponding to the energy of the incoming alpha. For
a given alpha energy we then searched for the value of the applied
reverse bias at which the saturation of the pulse height signals is
reached (see Fig.~\ref{fig04}-top panel). The correlation between
the evaluated range and the applied reverse voltage, shown in
Fig.~\ref{fig05}, is nicely reproduced by a square-root law
\cite{Kno00}.\\
%%%%%%%%%%%%%%%%%%%%%%%%%%%%%%%%%%%%
%%%%%%%%%%%%%%%%%%%%%%%%%%%%%%%%%%%%
%%%%%%%%%%%%%%%%%%%%%%%%%%%%%%%%%%%%
\begin{figure}[th]
\centering{\resizebox{0.6\textwidth}{!}{
\includegraphics{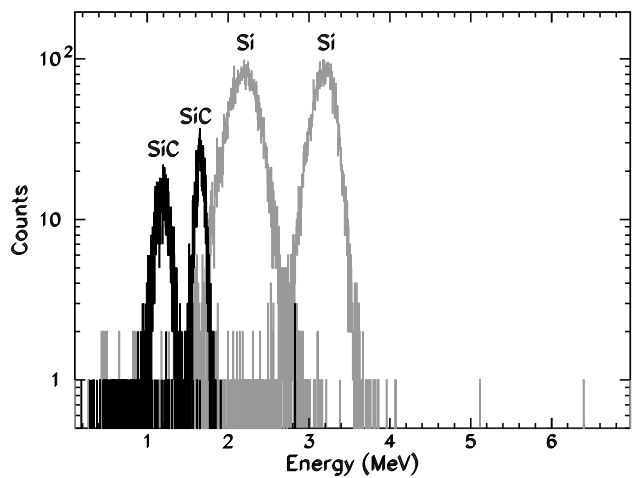}}}
\caption{Energy spectra of alpha particles of 2.19 and 3.18 MeV
incident energy measured in the silicon (gray) and in the SiC
(black) detectors. For comparison the energy values of the SiC
spectra were evaluated by using the energy calibration obtained for
the silicon detector.} \label{fig06}
\end{figure}
Finally, by correlating the saturated pulse height peak values and
the four alpha energies (2.08, 2.62, 3.09, 3.68 MeV) corrected for
the energy lost in the Ni$_2$Si 0.2 $\mu$m thick front layer of the
SiC diode, we calibrated the pulse height scale. By comparing the
energy peaks obtained with the SiC and the silicon detector for the
same deposited energy (see Fig.~\ref{fig06}) we have obtained a
factor of 2.1$\pm$0.3 in the charge produced by ionization, as
expected from the different ionization energy values of 3.76 eV in
Si and of 7.74 eV, as recently measured~\cite{Bert03,Ivan04}, in
SiC. \\To increase the explored energy range we used beams of
$^{12}$C at 5.06 and 17.68 MeV and $^{16}$O at 7.3, 9.78, 12.18 and
14.21 MeV of incident energies provided by the Tandem accelerator of
the LNS. The SiC diodes were operated at a reverse bias of  -100V
which depletes the active region of the detector up to 5.4$\pm$0.5
$\mu$m, wide enough to completely stop all the ions except the
$^{12}$C at 17.68 MeV and the $^{16}$O at 14.21 MeV of incident
energy which, according to SRIM calculation, release around 8 and 13
MeV respectively in the detector. Fig.~\ref{fig07} shows the energy
spectra measured by the SiC diodes. The high degree of linearity
observed in the correlation between the pulse height and the energy
shown in Fig.~\ref{fig08} for SiC detectors, is consistent with
previous measurements~\cite{Nav99,Rud03,Cun03}, and indicates the
proportionality of the produced ionization charge to the deposited
energy.
\begin{figure}[th]
\centering{\resizebox{0.6\textwidth}{!}{
\includegraphics{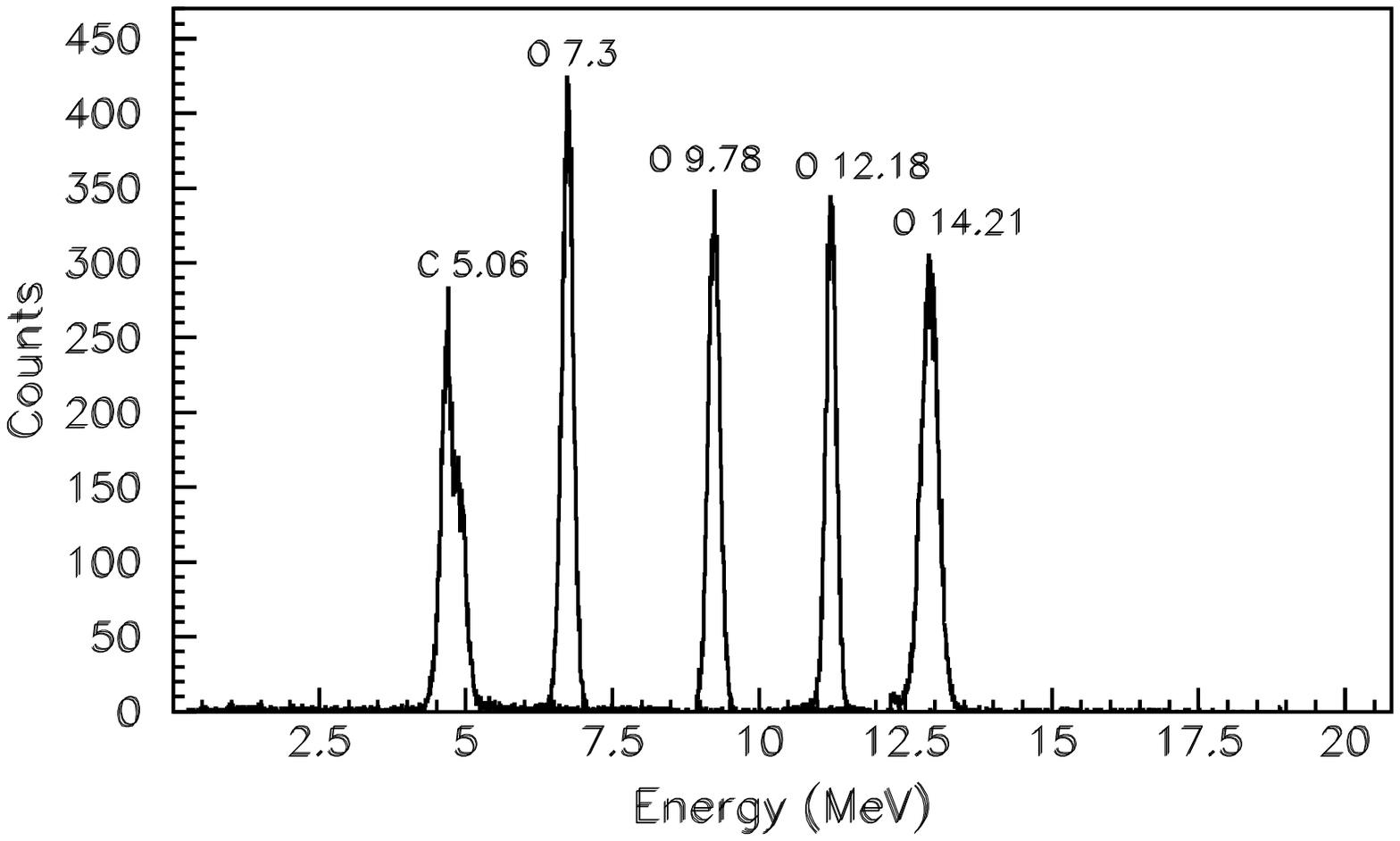}}
\resizebox{0.6\textwidth}{!}{
\includegraphics{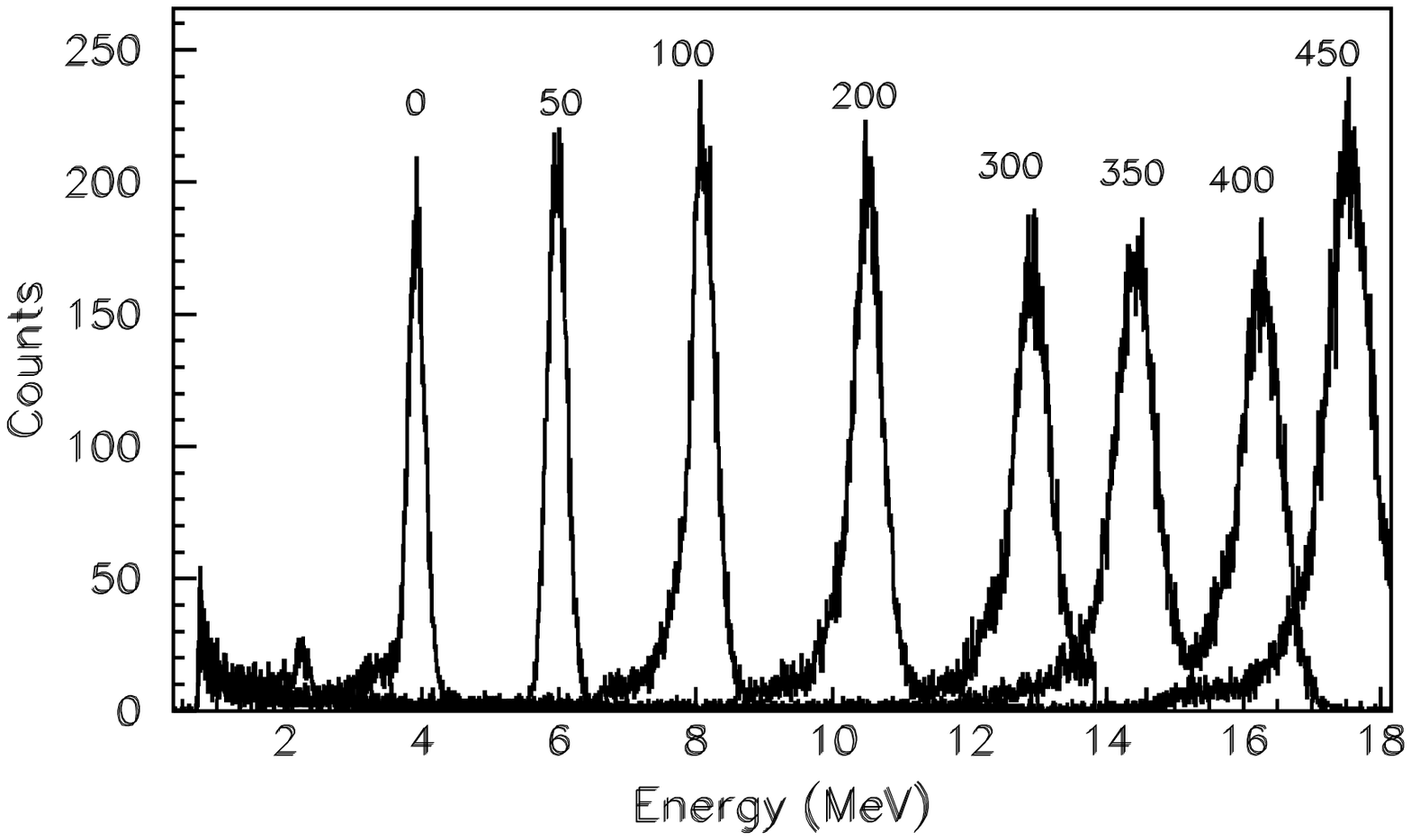}}}
\caption{Top: SiC energy spectra of $^{12}$C and $^{16}$O beams at
different energies. Bottom: 17.68 MeV $^{12}$C beam spectra measured
at increasing negative bias voltages.} \label{fig07}
\end{figure}
%%%%%%%%%%%%%%%%%%%%%%%%%%%%%%%%%%%%
Similarly to the measurements performed with the alpha source, a
number of spectra were taken over a range of bias voltages, from 0
to -~450 V, for the $^{12}$C beam at 17.68 MeV incident energy ( see
Fig.~\ref{fig07}-bottom panel).
\begin{figure}[th]
\centering{\resizebox{0.6\textwidth}{!}{
\includegraphics{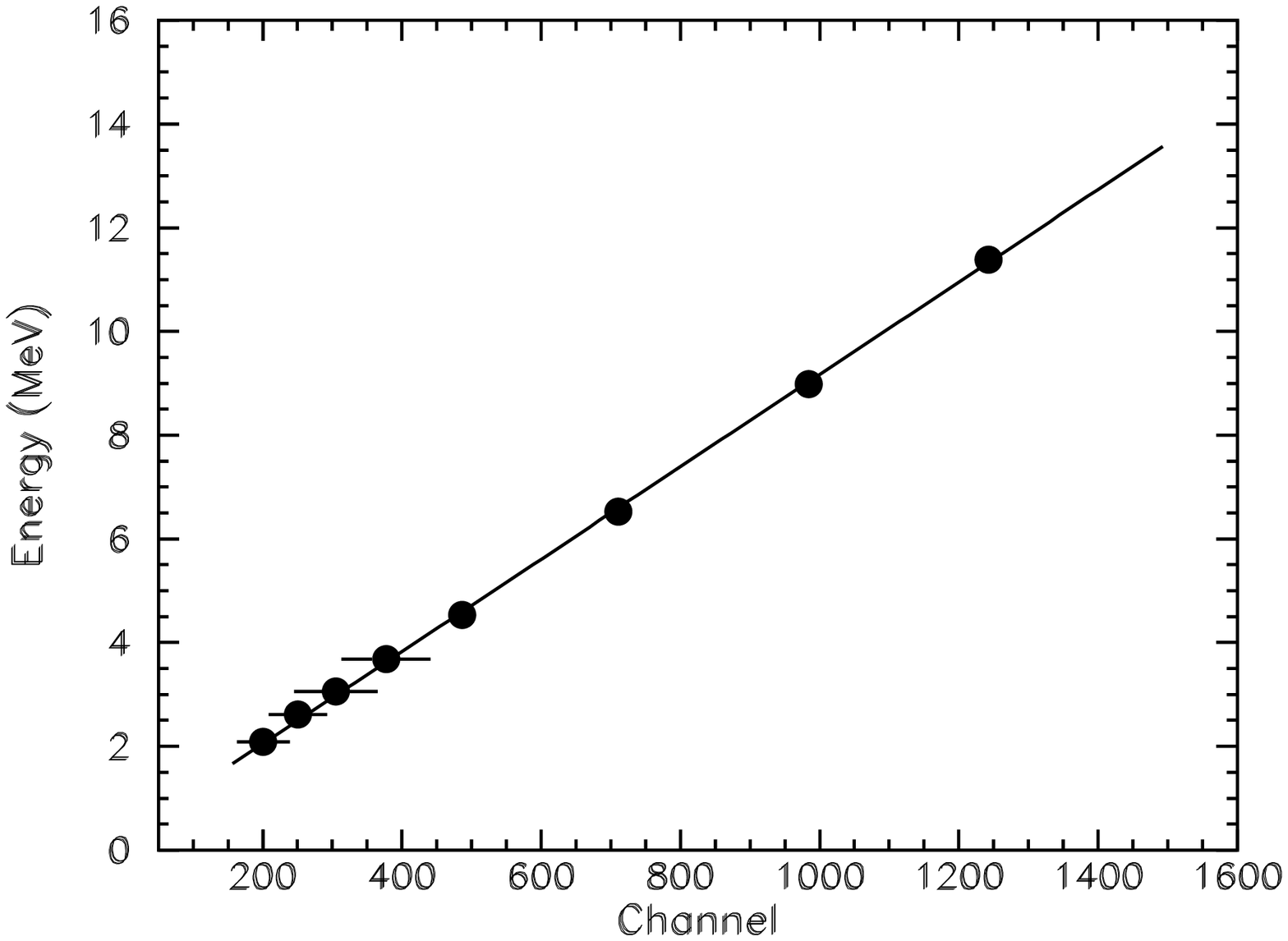}}}
\caption{Pulse height(channel) versus Energy plot. Data points refer
to the $^{12}$C (5.06 (4.5)MeV) and $^{16}$O (7.3 (6.5), 9.78 (9.0),
12.18 (11.4) MeV) beams and to the four alpha particles energies.
The energy values in parentheses are the ones corrected for the
energy lost in the Ni$_2$Si 0.2 $\mu$m thick front layer on the
detector. The linear correlation coefficient R$^2$ from the fit has
a value of 0.998.} \label{fig08}
\end{figure}
%%%%%%%%%%%%%%%%%%%%%%%%%%%%%%%%%%%%
We notice in Fig.~\ref{fig07} that no saturation has been observed
in the peak centroid values, since the active region of the detector
does not extend beyond the range of the $^{12}$C at 17.68 MeV in
SiC, which is approximately 11.2 $\mu$m. This occurrence indicates
that at the maximum applied bias voltage, -~450 V, the detector is
only about half depleted but, unfortunately, we could not raise the
bias voltage to higher values because the electrical connections in
vacuum were sparking at around -500 V.
 Therefore we evaluated the thickness of the depleted region by
comparing SRIM calculations to the energy loss measured at different
applied biases. Figure~\ref{fig09} shows the correlation between the
estimated thickness of the depletion layer and the applied reverse
bias for $^{12}$C ions. We remark the good agreement with the data
measured with alphas and the good linear correlation between the
depletion layer thickness and the square-root of the applied bias
values. From a linear fit (see Fig.~\ref{fig09}), at zero bias
voltage we estimated an active region thickness of 1.18 $\mu$m
resulting only from the Schottky contact potential, in good
agreement with the result of
Ref.~\cite{Nav99}.\\
\begin{figure}[h]
\centering{\resizebox{0.6\textwidth}{!}{
\includegraphics{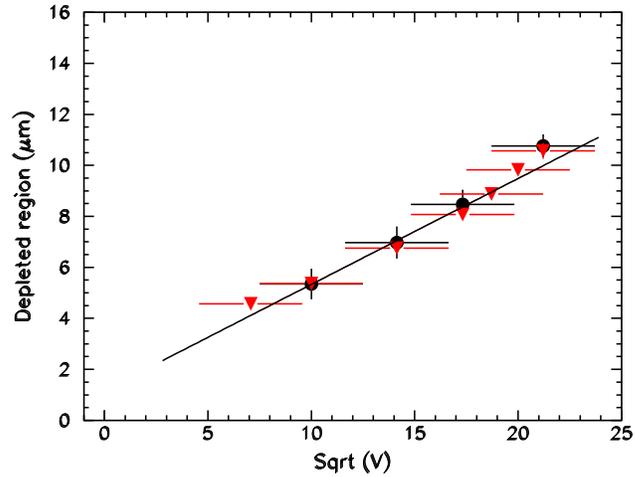}}}
\caption{Correlation between the square root of the applied bias and
the SiC depletion layer thickness evaluated from alpha (circles) and
the 17.68 MeV $^{12}$C beam (triangles) spectra. Error bars are
calculated from the FWHM of the energy loss spectra of
Fig.~\ref{fig07}.} \label{fig09}
\end{figure}
%%%%%%%%%%%%%%%%%%%%%%%%%%%%%%%%%%%%
In order to measure how the energy resolution behaves as a function
of the thickness of the active region and therefore of the applied
bias voltage, we analyzed the data reported in Fig.~\ref{fig07}. A
typical $^{12}$C 17.68 MeV response spectrum is shown in
Fig.~\ref{fig10}: the shape of the peak is well reproduced by a
Gaussian function.
\begin{figure}[ht]
\centering{\resizebox{0.6\textwidth}{!}{
\includegraphics{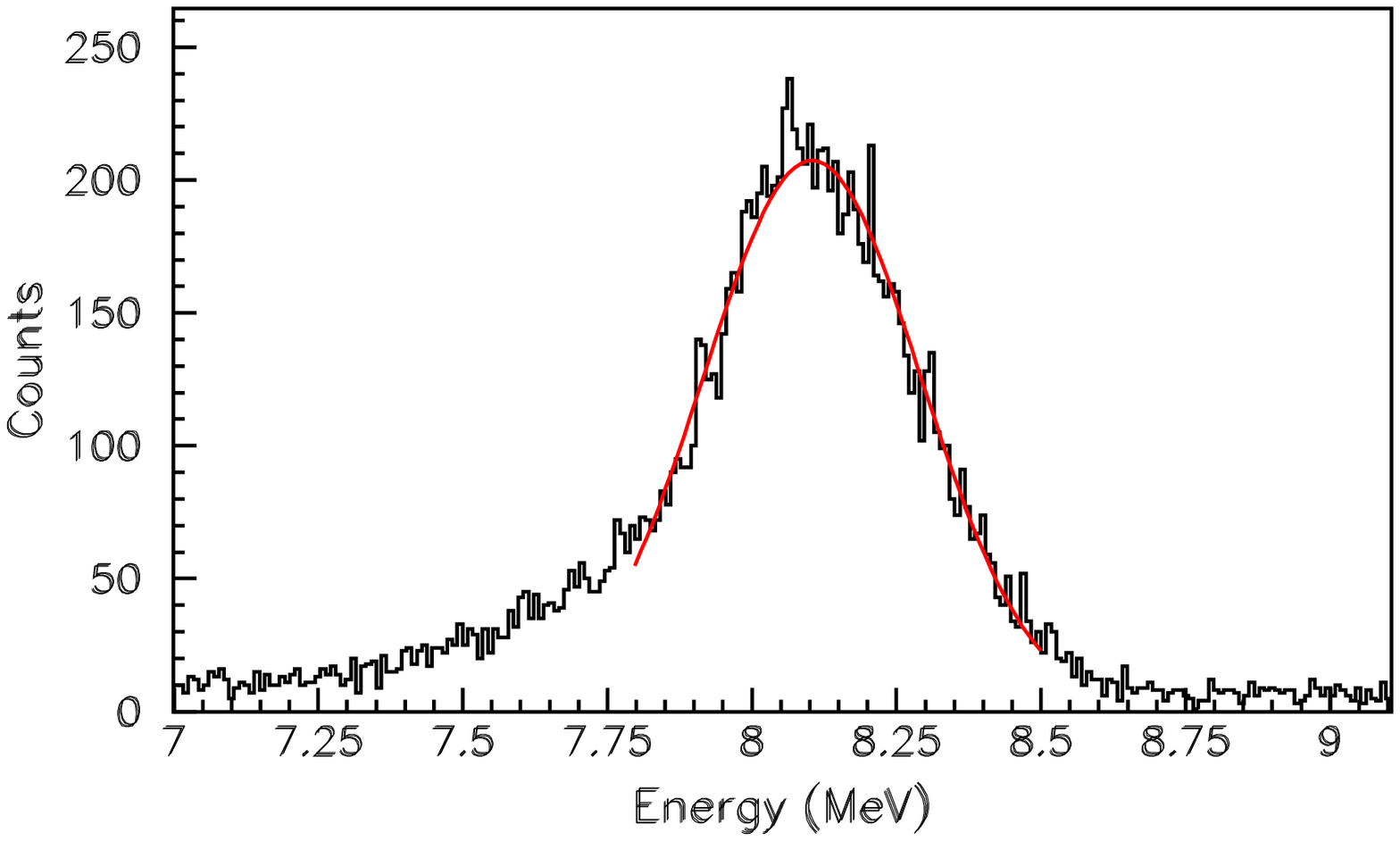}}
\resizebox{0.6\textwidth}{!}{
\includegraphics{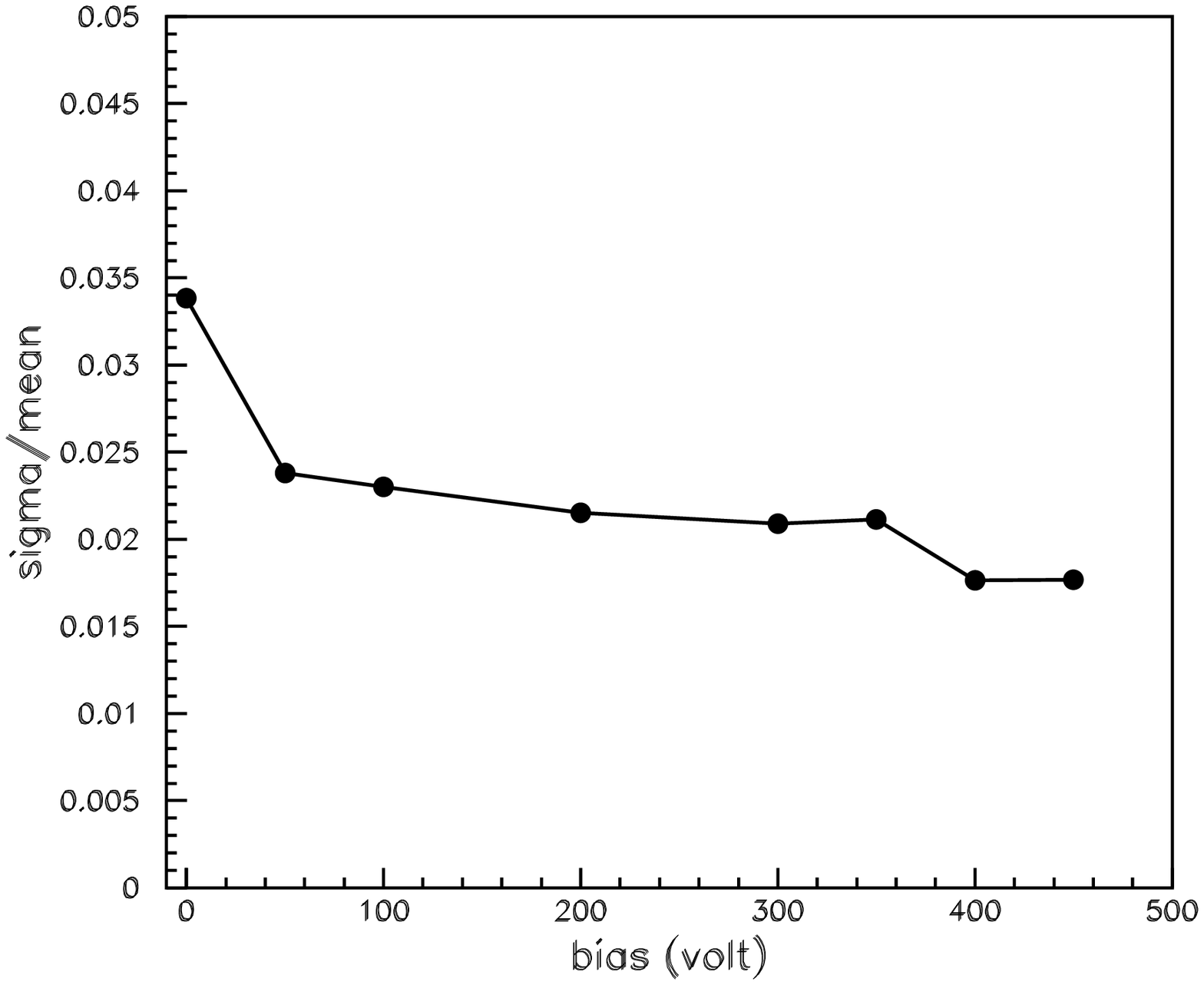}}}
\caption{Top: $^{12}$C at 17.68 MeV beam response of a SiC diode
with an applied bias voltage of -100 Volts. The spectrum is fitted
with a Gaussian function. Bottom: Relative energy resolution as a
function of the applied bias voltage.} \label{fig10}
\end{figure}
%%%%%%%%%%%%%%%%%%%%%%%%%%%%%%%%%%%%
Based on the energy calibration, the full-width at half-maximum
(FWHM) of the Gaussian fit to the $^{12}$C peak at 8.1 MeV (which is
the energy lost in 5.4 $\mu$m by the $^{12}$C at 17.68 MeV) is 190
keV corresponding to an energy resolution of 2.3\%. In the bottom
panel of Fig.~\ref{fig10}, the relative energy resolution (FWHM of
the peak over the peak position), is shown as a function of the
applied bias voltage. Besides an initial enhancement from 0 to -50
V, the relative resolution decreases slowly over the range of the
applied bias voltages up to a value of 1.8\%. The measured energy
resolution values are in good agreement with the ones observed in
other measurements~\cite{Rud03} although larger than the recently
reported SiC energy resolution value of 0.34\% ~\cite{Ivan04,Rud06}.

%%%%%%%%%%%%%%%%%%%%%%%%%%%%%%%%%%%%
\begin{figure}[h]
\centering{\resizebox{0.6\textwidth}{!}{
\includegraphics{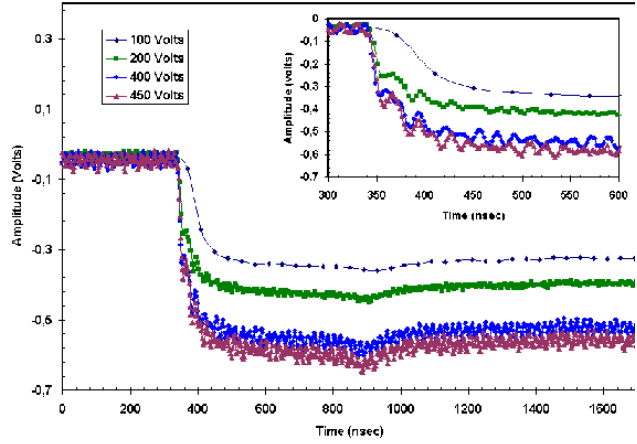}}}
\caption{Signals from the preamplifier generated by $^{12}$C ions at
17.68 MeV incident energy for different applied bias values.}
\label{fig11}
\end{figure}
%
%%%%%%%%%%%%%%%%%%%%%%%%%%%%%%%%%%%%

Moreover by storing the waveforms of the preamplifier output in the
Tektronix TDS 5104B Digital Oscilloscope we analyzed the behavior of
the pulses generated by the $^{12}$C beam at 17.68 MeV. Typical
pulses are reported in Fig.~\ref{fig11} for four different reverse
bias values. The amplitude variation is obviously related to the
increasing thickness of the active region and therefore to the
increasing energy lost by the $^{12}$C ions. The decrease of the
signal rise-time is related to the combined variations of the drift
velocity of the ionization charges, proportional to V$_{Bias}$, and
of the depleted region thickness, proportional to
$\sqrt{V_{Bias}}$.\\ Indeed, the rise-time varies from 97 nsec at
-100 Volts to 68 nsec at -200 Volts and reaches 46 and 44 nsec at
-400 and -450 Volts respectively. As expected the ratios of the
measured rise-time values scale as the inverse ratio of the
square-root of the applied bias voltages.\\ The consistency of the
SiC detectors performances were investigated by testing
charged-particle response of all the eleven detectors of identical
configuration. All of them showed, within the error bars, the same
characteristics here reported.

\section{Radiation damage}
\begin{figure}[h]
\centering{\resizebox{0.6\textwidth}{!}{
\includegraphics{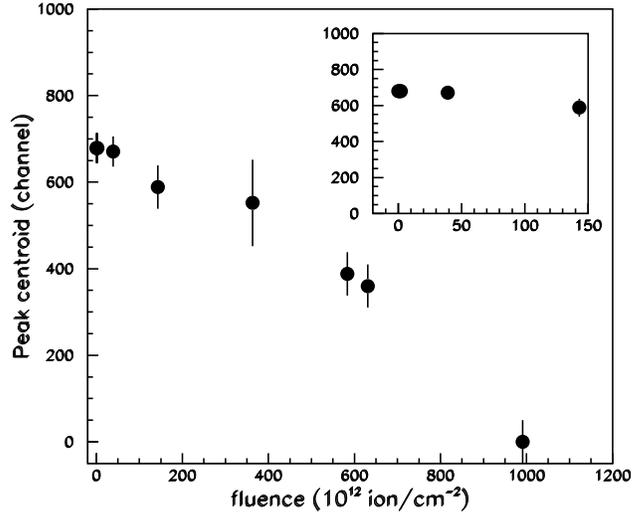}}}
\caption{Peak value of the spectrum of 53 MeV $^{16}$O beam measured
at increasing fluence. The plot at the lowest values of fluence is
expanded in the insert.}\label{fig12}
\end{figure}

\begin{figure}[h]
\centering{\resizebox{0.6\textwidth}{!}{
\includegraphics{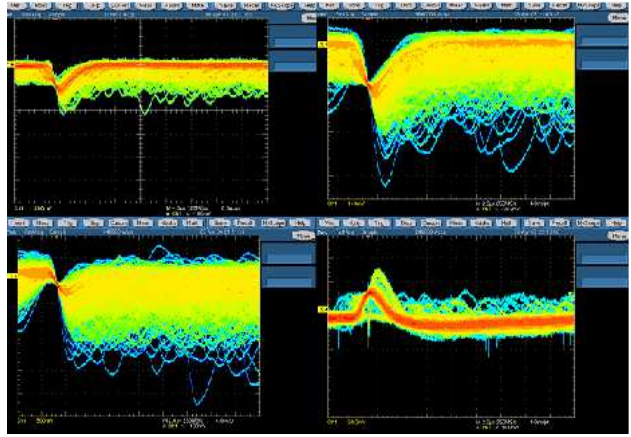}}}
\caption{Pulses from a SiC diode irradiated with the 53 MeV $^{16}$O
beam.
 Top left: beginning of irradiation. Top right: after a fluence of
3.5 10$^{14}$ ions/cm$^2$. Bottom left: after a fluence of 7
10$^{14}$ ions/cm$^2$. Bottom right: after a fluence of 10$^{15}$
ions/cm$^2$: notice the inversion of the signal.} \label{fig13}
\end{figure}

%%%%%%%%%%%%%%%%%%%%%%%%%%%%%%%%%%%%
One of the main motivations of such study was to explore the
radiation hardness properties of such SiC detectors. Five samples
were irradiated using 53~MeV $^{16}$O ions provided by the Tandem at
the LNS-Catania and their output signals were monitored during the
irradiation time. Fig.~\ref{fig12} shows the peak centroid of the
$^{16}$O spectrum as a function of the fluence.
%%%%%%%%%%%%%%%%%%%%%%%%%%%%%%%%%%%%
The applied reverse bias was kept fix at -100 V but the signals
appeared to decrease in amplitude as the irradiation increased. The
amplitude dropped to 50\% at a fluence of 6.5 10$^{14}$ ions/cm$^2$
(see Fig.~\ref{fig12}) in good agreement with results of
Ref.~\cite{Sci05,Qui04}. At the same time, the reverse current
increased by a factor of five and the noise by a factor of two (see
Fig.~\ref{fig13}). The present SiC diodes are therefore around ten
times radiation harder than silicon diodes but around a factor ten
less hard than thin epitaxial Si-diodes for which the charge
collection efficiency drops only to about 80\% at 6 10$^{15}$
protons/cm$^2$ at 24 GeV of incident energy \cite{Lin06}.

However it is not clear how to compare the damage produced by low
energetic $^{12}$C or $^{16}$O ions to the damage produced by
protons, neutrons, gammas or pions and which kind of defects, among
the known ones, are produced during irradiation.\\ Finally, two
different detectors were continuously observed during the
irradiation up to when they break-down at a fluence of 10$^{15}$
ions/cm$^2$. At this irradiation fluence we observed an inversion of
the signals with respect to the non-irradiated ones (see
Fig.~\ref{fig13}: bottom-right panel). This effect seems similar to
the symmetric Charge Collection Efficiency (CCE) response at both
polarities mentioned in Ref.\cite{Sci05} at a
fluence of 10$^{15}$ protons/cm$^2$ and attributed to radiation damage.\\
Moreover, we inverted both bias and amplifier polarities and we sent
on the irradiated detectors an $^{16}$O beam at 12 MeV incident
energy. Since the applied reverse bias of -100 V depletes 5.4 $\mu$m
of the SiC active region, we expected an energy release of 8.3 and
12 MeV in the detector from the $^{16}$O beam at 53 and 12 MeV
incident energy respectively, but the pulse height of the signal was
the same in both cases. Therefore we conclude that the detector was
sensitive to the charged ion but not to the deposited energy.\\
Further analysis will be performed to understand if the observed
effect is of the same nature of space-charge sign-inversion
(``type inversion''), reported for
Si-diodes~\cite{Lin01,Ano02,Bos03}. At the moment we suggest, as
possible explanation, the formation of a layer of $^{16}$O
(10$^{15}$ ions/cm$^2$) at a distance of 27.2$\pm$0.4 $\mu$m
(which is the range of the 53 MeV $^{16}$O ions in the SiC
material) from the Schottky contact, and therefore in the n$^+$
type substrate, subdividing the 300$\mu$m thick material in two
parts with a floating ground at the $^{16}$O layer.

\section{Conclusions}

In the present work we investigated the response of SiC Schottky
diodes to alphas, $^{12}$C and $^{16}$O low energetic ions in order
to explore the possibility of using these detectors in nuclear
physics applications in extreme environments. The signal response to
the ionization produced by the low-energy ions was analyzed in terms
of linearity, energy resolution, rise-time and deterioration as
function of the applied reverse bias and the irradiating fluence.
The latter measurements demonstrate the good quality of the SiC as a
radiation hard material. \\Energy resolution on the order of 3\% and
rise time up to 44 nsec have been measured at the maximum applied
reverse bias. The amplitude of the signal drops down to 50 \% at a
fluence of 6.5 10$^{14}$ ions/cm$^2$ indicating more than an order
of magnitude hardness to the radiation with respect to silicon
diodes. The signal inversion at a fluence of 10$^{15}$ ions/cm$^2$
should be further investigated to understand if it is produced by
radiation damage or by the $^{16}$O implant in the material. \\The
main inconvenience we have found in the use of the present detectors
was the high bias voltage, needed to deplete the active region of
the diode. We expect a correlation between the depleting bias values
and the N doping concentration in the epitaxial region \cite{Nav99}.
We will investigate in the near future such a correlation, being
aware of the negative influence on the ionization charge production
and radiation hardness of a reduced doping concentration. But
reduction of the applied voltage to values far from the break down
ones is a very crucial issue for future applications.

\section{Acknowledgements}
The help in the SiC diodes selection and assembling and the valuable
discussion with L.Calcagno, G.Foti and F.La Via are greatly
appreciated. \ E.R. acknowledges the support from the European
Community in the framework of the "DIRAC secondary-beams" -contract
N.515873 under the "Structuring the European Research Area" Specific
Programme Research Infrastructures action.

%%%%%%%%%%%%%%%%%%%%%%%%%%%%%%%%%%%%%%%%%%%%%%%%%%%%%%%%
% End of the paper
%

%------------------------------ BIBLIOGRAPHY

\end{document}